\documentclass[letterpaper, 10 pt, conference]{ieeeconf}  
\makeatletter
    \let\NAT@parse\undefined
\makeatother

\IEEEoverridecommandlockouts                              

\usepackage{graphics} 
\usepackage{amsmath} 
\usepackage{amssymb}  
\usepackage{todonotes} 

\let\labelindent\relax
\usepackage{enumitem}
\usepackage{flushend}
\usepackage[numbers,sort&compress]{natbib}
\usepackage[colorlinks = true,
            linkcolor = black,
            urlcolor  = blue,
            citecolor = magenta,
            breaklinks = true,
            bookmarks = true, pagebackref]{hyperref}

\title{\LARGE \bf
OPTIMA: Optimized Policy for Intelligent Multi-Agent Systems Enables Coordination-Aware Autonomous Vehicles
}

\author{Rui Du$^{1}$, Kai Zhao$^{1}$, Jinlong Hou$^{1}$, Qiang Zhang$^{1}$ and Peter Zhang$^{2}$
\thanks{$^{1}$Rui Du, Kai Zhao, Jinlong Hou, Qiang Zhang are with the AI Platform Department, Bilibili Inc., Shanghai 200233, China
        {\tt\small durui@bilibili.com; zhaokai02@bilibili.com; houjinlong@bilibili.com;zhangqiang@bilibili.com} 
        }%
\thanks{$^{2}$Peter Zhang is with the Heinz College of Information Systems and Public Policy, Carnegie Mellon University, 4800 Forbes Ave, Pittsburgh, PA 15213, USA,
        {\tt\small pyzhang@cmu.edu}}%
}

\begin{document}

\maketitle
\thispagestyle{empty}
\pagestyle{empty}

\begin{abstract}
    Coordination among connected and autonomous vehicles (CAVs) is advancing due to developments in control and communication technologies. However, much of the current work is based on oversimplified and unrealistic task-specific assumptions, which may introduce vulnerabilities. This is critical because CAVs not only interact with their environment but are also integral parts of it. Insufficient exploration can result in policies that carry latent risks, highlighting the need for methods that explore the environment both extensively and efficiently. This work introduces OPTIMA, a novel distributed reinforcement learning framework for cooperative autonomous vehicle tasks. OPTIMA alternates between thorough data sampling from environmental interactions and multi-agent reinforcement learning algorithms to optimize CAV cooperation, emphasizing both safety and efficiency. Our goal is to improve the generality and performance of CAVs in highly complex and crowded scenarios. Furthermore, the industrial-scale distributed training system easily adapts to different algorithms, reward functions, and strategies.
\end{abstract}


\section{INTRODUCTION}
The long-term goal of autonomous vehicles is to address advanced real-world traffic challenges. According to a report by the National Highway Traffic Safety Administration (NHTSA) \cite{national2008national}, 94\% of all crashes are caused by human error. However, human-error crashes stem not only from impaired or distracted driving but also from misjudging other vehicles' intentions. The key to eliminating crashes lies in the connectivity of vehicles and their ability to achieve comprehensive situational awareness, enabling them to respond appropriately to each other's movements and decisions. Real-world traffic problems are complex and unpredictable. Thus, training a highly intelligent model that understands complex roads and diverse driving styles is necessary.

However, the majority of previous studies have focused on overly simplistic scenarios in simulations. This approach may lead to potential risks affecting the overall safety and effectiveness of autonomous driving technologies. CAVs are not only interacting with the environment they are also integral parts of it. Each CAV's actions influence the surroundings, which in turn affects other CAVs, creating a complex feedback loop that simplistic simulations often fail to capture. If CAVs are not exposed to a wide range of scenarios during training, they may be ill-prepared for real-world driving diversity. When encountering unfamiliar situations, CAVs might exhibit unexpected behaviors, potentially triggering a cascade effect where other vehicles are pushed into their own corner cases. This interconnectedness underscores the need for comprehensive, diverse, and challenging simulations to develop robust and safe CAV systems capable of handling the unpredictable nature of real-world traffic.



This paper seeks to enhance the scalability of autonomous driving systems by utilizing distributed reinforcement learning techniques. We aim to address traditional traffic problems by integrating advanced AI models that can efficiently process and react to complex traffic scenarios with enhanced safety. The main contribution of our work is a novel \textbf{O}ptimized \textbf{P}olicy for In\textbf{t}ell\textbf{i}gent \textbf{M}ulti-\textbf{A}gent Systems, or OPTIMA. Our key contributions are summarized as follows:


\begin{itemize}
    \item We have integrated learning-based methods with established perception and cooperation techniques such as centralized policy, safety distances, right-of-way.
    \item We have successful implemented distributed reinforcement learning for autonomous vehicles, enhancing scalability and performance in complex scenarios.
    \item OPTIMA achieved a 100\% success rate in navigating extremely challenging multi-agent cooperation scenarios, a feat previously unattained at this level of environmental complexity.
\end{itemize}
By addressing these complex scenarios without relying on simplifying assumptions, OPTIMA sets a new benchmark for autonomous driving research. It demonstrates the potential of distributed reinforcement learning in handling real-world traffic complexities and provides valuable insights for future research and development in this field.

\section{Related Work}

\begin{figure*}[ht]
    \centering
    \includegraphics[width=\textwidth]{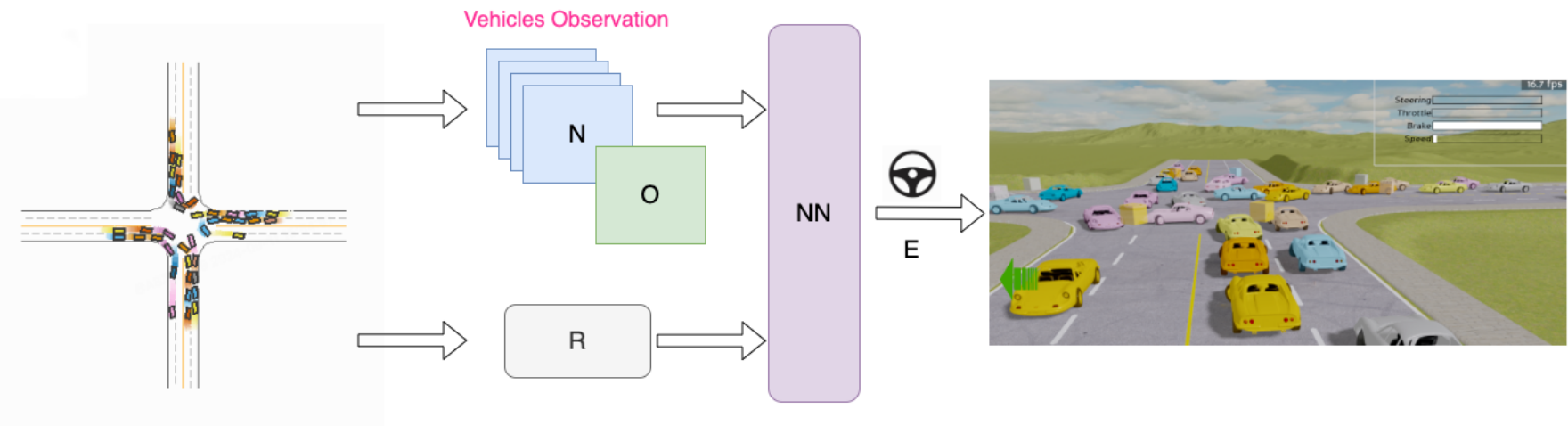}
    \caption{The process where the model receives observations about a vehicle and its neighboring vehicles, denoted as \( O \) for the vehicle's own observation and \( N \) for the neighboring vehicles' observations, from the environment. Using a deep reinforcement learning model, represented as \( NN \), it generates appropriate actions to control the vehicle's response and maneuvers. The model also involves a reward function \( R \), which influences the actions based on predefined criteria. \( NN \) outputs not only the control actions but also an estimation of future outcomes, represented as \( E \).
}
    \label{fig:architecture}
    \vspace*{-4mm}
\end{figure*}

To define what constitutes safety and efficiency in traffic, we have reviewed a variety of literature on the subject. Some studies have explored these aspects by focusing on varied hazardous scenarios and interaction dynamics between autonomous and human-driven vehicles. For instance, one study evaluated and compared 33 base metrics and 51 variants of traffic safety indicators published from 1967 to 2022~\cite{singh2023diversity}. Another paper discusses the safety of autonomous vehicles with great care, proposing a white-box, interpretable mathematical model for safety assurance, which the authors call Responsibility-Sensitive Safety (RSS)~\cite{shalev2017formal}. Other papers discuss various safety indicators and challenges across different scenarios in detail~\cite{abbas2023autonomous, shetty2021safety, zhang2024analysis, wang2020many}. Inspired by these works, we have determined that vehicle safety indicators can be defined by safe distances, intent recognition, field of view limitations, and the safety of sensitive areas. After summarizing these issues and challenges, we factorize the exploration problem into (i) learning to understand the intentions of other vehicles and perceive surrounding environmental information, and (ii) making appropriate operational decisions based on this understanding.

Several works have proposed learning-based approaches to enhance the perception of surrounding vehicles in CAVs context. A particular study talks about two primary decision-making strategies: pipeline planning and end-to-end planning~\cite{teng2023motion}. Pipeline planning, often referred to as rule-based planning, is a traditional approach the planning within a broader system that includes perception, localization, and control~\cite{song2022msfanet, gong2022real}. This method forms a critical part of the overall framework necessary for executing autonomous driving functions. Conversely, end-to-end planning represents a more holistic approach where the entire driving function from perception to action is encapsulated within a single, comprehensive model.

Moreover, research into end-to-end reinforcement learning methods is increasingly being explored, as exemplified by studies such as~\cite{valiente2023prediction, wang2021harmonious, kiran2021deep, li2022decision, zhang2022multi, li2022metadrive, lyu2021contrasting, peng2021learning, han2020multi}. These approaches are promising for their ability to directly map sensory inputs to driving actions, potentially simplifying the complex process of autonomous decision-making. However, these methods often remain confined to simplistic scenarios and struggle with scalability.

Existing RL frameworks each have their own limitations. For instance, the CleanRL framework, designed with a single-file structure for research-friendly features, prioritizes ease of learning over scalability~\cite{huang2022cleanrl}. Stable Baselines3, while offering reliable RL algorithms, lacks support for asynchronous multi-actor parallel capabilities~\cite{raffin2021stable}. Ray RLlib, despite its powerful features and multi-machine scalability, has a steep learning curve and deeply nested abstractions that can hinder customization for specific tasks~\cite{liang2017ray, liang2018rllib}. These limitations become particularly apparent in multi-agent reinforcement learning (MARL) cooperation tasks like intersection, which differ significantly from traditional zero-sum competitive tasks such as Go~\cite{lv2018approximate}. Cooperative tasks often require a delicate balance between team and individual rewards, a nuance not easily captured in existing frameworks. Furthermore, the partial observation of agents in traffic scenarios adds another layer of complexity. Adapting existing open-source frameworks to meet these specific needs often requires extensive code modifications, which can limit their flexibility for complex, cooperative traffic tasks. This highlights the need for a more adaptable and scalable framework designed specifically for complex MARL scenarios in autonomous driving. OPTIMA enhances this approach by incorporating distributable, asynchronous actors, which facilitate scalability and are well-suited for traffic cooperation scenarios.

\begin{table}
\caption{Summary of RL Framework}
\label{table_rl_algorithms_perf}
\begin{center}
\begin{tabular}{p{1.5cm} p{1.5cm} p{0.9cm} p{0.9cm} p{1.2cm}}
\hline
\textbf{Framework} & \textbf{Distributable} & \textbf{Async Actors} & \textbf{Scalable} & \textbf{Traffic Solutions} \\
\hline
CleanRL            &  $\times$                & $\times$               & $\times$           & $\times$        \\
SB3                & $\times$                 & $\times$               & $\times$           & $\times$        \\
Ray RLlib          & $\checkmark$             & $\checkmark$           & $\times$           & $\times$        \\
OPTIMA             & $\checkmark$             & $\checkmark$           & $\checkmark$       & $\checkmark$    \\
\hline
\end{tabular}
\end{center}
\end{table}


\section{Preliminary}

\subsection{Markov Decision Process Formulation}

We formulate the task as a set of Decentralized Partially Observable Markov Decision Processes (Dec-POMDPs)~\cite{6760239}. To accommodate the evolving vehicles in the complex scenarios, Dec-POMDPs are represented as a tuple \( \langle S, \{A_i\}, T, R, \{\Omega_i\}, O, \{N_i\}, \gamma \rangle \). We defined the environmental time step of each agent as \( t_i \in T \). \( S \) is a set of global states. At each time step \( t \), the action chosen by an agent is \( a_{i,t} \in A_{i,t} \), where \( A_{i,t} \) is a two-dimensional continuous action space in the environment. Each agent receives its own observation at each time step \( t \), denoted as \( s_{t, i} \), which includes a range of sensor inputs such as lidar data, vehicle dynamics, and lane information. Additionally, each agent \( i \) has a set of neighbors \( N_i \).
The observation function \( O \), defined by \( O(s', a, o) = P(o \mid s', a) \) represents the set of conditional observation probabilities. \( \gamma \in [0,1] \) is the discount factor for reward function \( R \). 
The reward function \( R \) consists of the distance and speed driven on the road and the final arrival at the destination. Of course, if there is a collision or driving off the road, there will be corresponding penalties.

The action space \(A\) is designed to accommodate the dynamics of each of the 40 agents within the environment. At any given time step \( t \), the action \( a_{t,i} \) for each agent \( i \) is derived from a specific set of control parameters indexed by \( j \). The composite action vector for each agent, denoted \( a_{t,i} = (a_{t,i,0}, a_{t,i,1})\), composes
\begin{itemize}
    \item \( a_{t,i,0} \in [-1, 1] \): Steering angle, with -1 and 1 indicating the maximum left and right turns, respectively.
    \item \( a_{t,i,1} \in [-1, 1] \): Throttle and brake control, where positive values signify forward acceleration, negative values denote braking, and reverse movement is enabled if the vehicle’s speed is less than or equal to zero.
\end{itemize} 

\section{Methods}

\begin{table}
\caption{Performance of RL Algorithms}
\label{table_rl_algorithms_perf}
\begin{center}
\begin{tabular}{lrrr}
\hline
                                     &     DDPG &     SAC &    PPO \\
\hline
 success                             &     3.31 &   10.49 &  \textbf{40}    \\
 out\_of\_road                         & 26061.2  &   69.14 &  \textbf{14.39} \\
 crash\_vehicle                       & 16542.9  &  171.31 &  \textbf{31.43} \\
 velocity\_mean                       &     1.87 &    0.82 &   \textbf{4.11} \\
 episode\_steps                       &  1000    & 1000    & \textbf{478.92} \\
\hline
\end{tabular}
\end{center}
\end{table}

In the complex landscape of urban transportation, intersections stand out as critical environment, often serving as the primary source of traffic congestion. Within the framework of intelligent transportation systems, the coordinated management of intersection traffic emerges as a crucial component. This approach leverages vehicle-to-infrastructure and vehicle-to-vehicle communication capabilities, offering promising avenues for enhancing both road safety and traffic efficiency. Our study focuses on a particularly challenging scenario: the coordination of autonomous vehicles at a four-way intersection without traditional traffic signals.
\subsection{Experiment Settings}


\begin{table*}[ht]
\caption{Experimental Results: Impact of Safe Distance and Right-of-Way Rules}
\label{table:safety_result}
\begin{center}
\begin{tabular}{lrrrr}
\hline
                                     &   baseline &   safe\_distance &   right\--of\--way &   safe\_distance \&  right\--of\--way \\
\hline
 success                             &      40    &           40    &          40    &                              40    \\
 out\_of\_road                         &      14.39 &           21.53 &          13.51 &                             \textbf{13.48} \\
 crash\_vehicle                       &      31.43 &           14.62 &          31.77 &                              \textbf{12.61} \\
 velocity\_mean                       &       \textbf{4.11} &            3.97 &           3.9  &                               3.97 \\
 velocity\_mean\_in\_conflict\_zone      &       2.47 &            3.01 &           2.23 &                               \textbf{3.31} \\
 acceleration                        &       0.6  &            0.55 &           \textbf{0.64} &                               0.59 \\
 acceleration\_in\_conflict\_zone       &       0.43 &            0.53 &           0.43 &                               \textbf{0.6}  \\
 arrive\_steps                        &     \textbf{277.96} &          285.46 &         295.67 &                             283.02 \\
 episode\_steps                       &     \textbf{478.92} &          498.19 &         504.84 &                             491.62 \\
 mean\_conflict\_zone\_num              &       5.73 &            4.34 &           6.4  &                               \textbf{3.91} \\
 max\_conflict\_zone\_num               &      12.12 &            8.72 &          13.02 &                               \textbf{8.2}  \\
 conflict\_zone\_when\_crash            &       8.34 &            5.86 &           8.9  &                               \textbf{5.25} \\
 front\_end\_distance                  &       0.18 &            \textbf{0.2}  &           0.17 &                               \textbf{0.2}  \\
 limited\_lidar                       &       0.56 &            0.64 &           0.55 &                               \textbf{0.67} \\
 limited\_lidar\_in\_conflict\_zone      &       0.43 &            0.53 &           0.41 &                               \textbf{0.56} \\
 front\_end\_distance\_in\_conflict\_zone &       0.12 &            0.15 &           0.11 &                               \textbf{0.16} \\
 pair\_distance                       &      37.76 &           41.82 &          36.4  &                              \textbf{45.24} \\
\hline
\end{tabular}
\end{center}
\end{table*}
\paragraph{Enhancing Simulation Complexity and Precision}
To ensure both the efficiency of training and the realism of the simulation environment, we selected  MetaDrive\footnote{MetaDrive can be found at: \raggedright\url{https://github.com/metadriverse/metadrive}}, a lightweight 3D traffic simulator optimized for the training and evaluation of MARL methods. MetaDrive is particularly suited for scalable deployment across distributed clusters, accommodating the increasing complexity of training tasks as the number of agents grows. This choice was driven by the need for extensive training data to support complex multi-agent cooperative tasks. Unlike other simulators that may offer richer visual effects or features~\cite{dosovitskiy2017carla, shah2018airsim, krajzewicz2010traffic}, MetaDrive is designed to be lightweight, ensuring easier deployment on Linux servers. It provides reinforcement learning friendly APIs, such as encapsulated reward functions and observation feature extraction, facilitating efficient training. In our study, we configured the simulator with 40 vehicles in an intersection scenario without traffic lights to assess agent cooperation. 


Many simulations setting where vehicles disappear after collisions, our setup maintains crashed vehicles on the road~\cite{peng2021learning}. This design choice significantly increases the complexity of the environment and more accurately reflects real-world traffic conditions. Removing crashed vehicles can inadvertently simplify the traffic flow, potentially leading to less realistic training scenarios. By keeping collided vehicles in place, we ensure that agents must learn to navigate around obstacles and deal with the ongoing consequences of accidents, just as they would in real-world driving situations. Moreover, to achieve more precise control over the vehicles, we use continuous action in simulation. Many studies in the field assume macro-level actions for vehicle control~\cite{han2020multi}. However, this assumption leads to several limitations, particularly in complex traffic scenarios. Macro-level actions often result in reduced adaptability and responsiveness, limiting the ability of autonomous vehicles to execute precise maneuvers necessary for safe and efficient driving. 

\paragraph{Reinforcement Learning Algorithms}
%

Reinforcement learning has demonstrated remarkable efficiency and effectiveness across a variety of domains~\cite{sutton2018reinforcement}, particularly in multi-agent settings\cite{lowe2017multi}. The abundance of available algorithms provides a rich toolkit for addressing complex tasks such as autonomous driving. In this study, our goal is to evaluate the performance of different RL algorithms in enhancing the safety aspects of autonomous driving and to identify which algorithm achieves superior experimental results.

Given the varied performance of different RL algorithms across diverse tasks, it is crucial to undertake a comparative analysis to discern their strengths and weaknesses in specific scenarios. We employed a distributed training setup utilizing 4 GPUs and 256 CPUs to train three prominent reinforcement learning algorithms: Deep Deterministic Policy Gradient (DDPG), Soft Actor-Critic (SAC), and Proximal Policy Optimization (PPO)~\cite{lillicrap2015continuous,haarnoja2018soft,schulman2017proximal}. Training was conducted over 24 hours, simulating approximately 250 million data instances. Due to the distributed nature of our simulation across various CPUs within the same cluster, data communication was asynchronous.

The results, summarized in the table below Table \ref{table_rl_algorithms_perf}, clearly demonstrate that PPO significantly outperformed DDPG and SAC in this task. Notably, PPO achieved a perfect success rate with all 40 vehicles passing the test, which was not matched by the other algorithms. Consequently, PPO will be referred to as the baseline algorithm in subsequent discussions within this paper.

\begin{figure}[h]
    \centering
    \includegraphics[width=\linewidth]{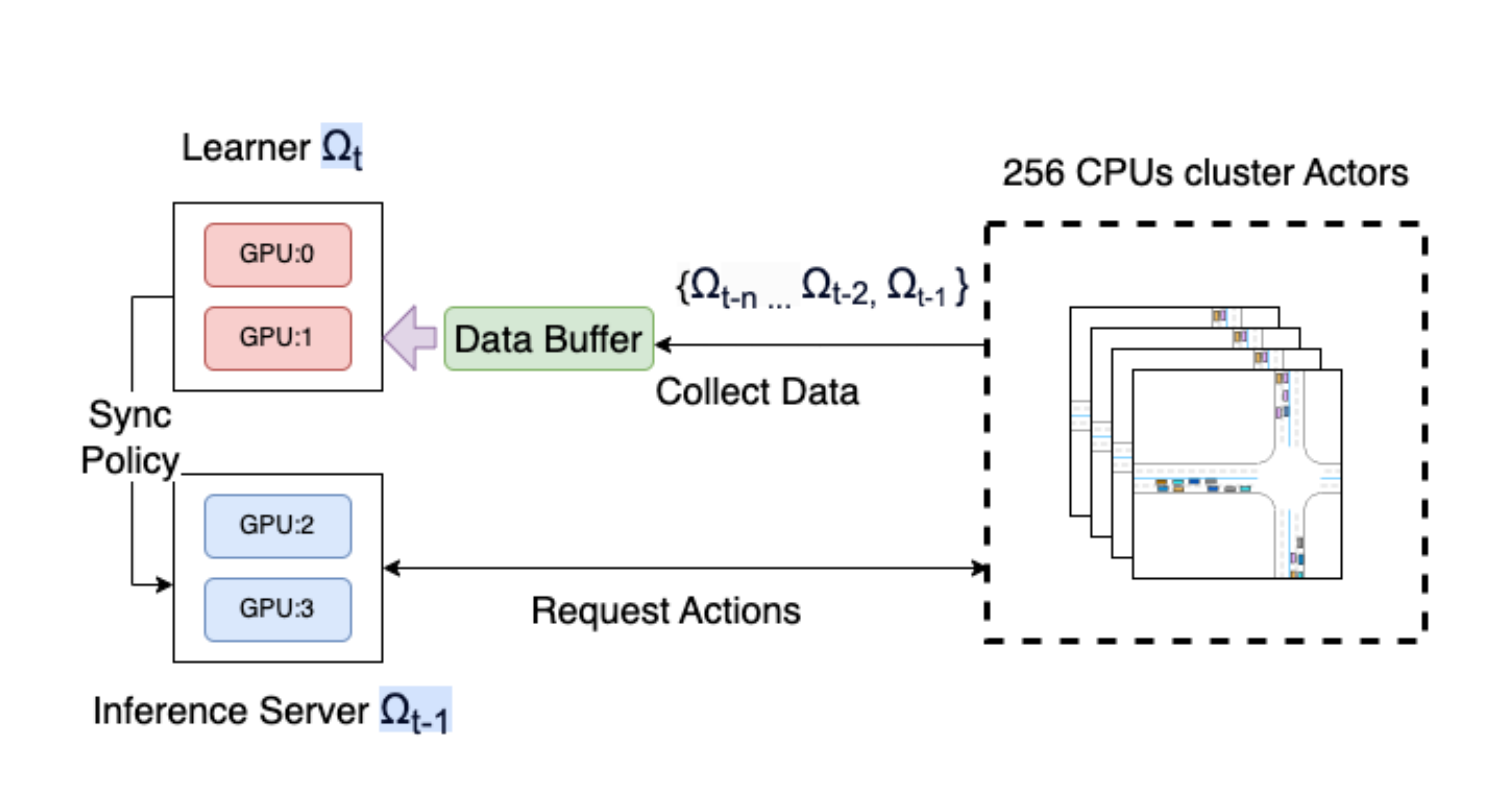}
    \caption{Architecture of the distributed training system.}
    \label{fig:mysys}
\end{figure}

\paragraph{Advanced Distributed Training for Collaborative Efficiency} To enhance the speed and efficiency of our experiments, we implemented a distributed training system as illustrated in Figure \ref{fig:mysys}. This system consists of 256 CPU cores and 4 NVIDIA V100 GPUs. 

In this setup, two GPUs were dedicated to inference tasks while the other two were used for training. This separation allowed us to optimize the parallel processing of data generation and model training, overcoming the limitations of single-machine setups where inference and training often occur sequentially, leading to bottlenecks. As a result, our system was able to generate approximately 15,000 data points per minute and perform 256 gradient descent updates per minute, significantly improving the efficiency of our reinforcement learning experiments, especially in these complex tasks.

Unlike zero-sum environments where rewards are strictly competitive, cooperative tasks in MARL require more nuanced reward distributions to prevent overfitting due to disproportionately large rewards for some agents. To address this, we using reward normalization techniques within our OPTIMA framework. Given that data is centrally collected in the data buffer, we can apply batch normalization or moving averages to rewards or advantages during sampling. This method is superior to local normalization as it allows for significantly larger batch sizes~\cite{schaul2021return}, enhancing the stability and effectiveness of our training procedures.

\paragraph{Asynchronous Optimization Trade-offs} While distributed reinforcement learning offers significant advantages in terms of scalability and efficiency, it also introduces certain challenges, particularly in the context of asynchronous optimization. The decoupling of various modules in the system, while beneficial for parallelization, can lead to version inconsistencies between the model being trained and the one used for inference. For instance, while the model is being updated during training $\Omega_t$, the actors may be simultaneously conducting inference using a slightly older version of the model $\Omega_{t-1}$ or even earlier $\Omega_{t-n}$. This asynchronous nature results in a situation where the collected training data is always a few versions behind the policy $\Omega_t$ currently being trained. This stale data can potentially impact the stability and convergence of the learning process, even caused the training to crash, requiring careful consideration in the design and implementation of the distributed system. 

Moreover, in practice, we cannot simply discard this stale data during the training process. There are two primary reasons for this. First, discarding data reduces data utilization efficiency, potentially leading to insufficient data for training, where the learner constantly waits for new data. Second, and more critically, in the actual training process, data is not sent to the learner from the actor until it reaches a certain Horizon length. However, this situation often occurs during the last few crucial steps of an episode, such as when a CAV reaches its destination, which are generally the most valuable learning experiences. Furthermore, due to the complex environment reset and warm-up times, the time span for collecting this crucial data is extended. Therefore, striking a balance between the amount of exploratory data for training and managing stale data becomes particularly crucial in this context.

To address these challenges, we propose a straightforward and universally applicable method that doesn't require modifying the loss function, making it suitable for any RL algorithm. Our approach involves managing stale data in the sampling data buffer, ensuring a certain level of data freshness. Specifically, in our system settings, we work with data batches of size 4096 and a horizon length of 128, resulting in a data set of 4096 x 128 elements. For this set, we set a maximum allowable average gap of 8 versions between the data $\Omega_{t-n}$ and the current Learner model $\Omega_t$. If this threshold is exceeded, we discard the older data from the queue. We calculated that this approach results in discarding approximately 13\% of the data, which we consider acceptable without significantly impacting the training effectiveness. This method strikes a balance between data utilization and training stability, addressing the challenges of asynchronous optimization in a practical and efficient manner.

\subsection{Policy Coordination: Decentralized vs Centralized}
Building upon our distributed training framework, we now turn our attention to a crucial aspect of multi-agent systems: policy coordination. To study the trade-off between efficiency and safety for decentralized intelligent vehicles and centralized interconnected intelligent vehicles, we consider two approaches:

Centralized Training with Decentralized Execution (CTDE): In standard CTDE~\cite{chen2019new}, agents are trained using global information but execute actions based only on local obsetions. This approach does not involve pooling of hidden variables during execution.

Centralized Training with Cecentralized Execution (CTCE): We implement a fully centralized method that extends beyond traditional CTDE. We use centralized communication and policy coordination through the pooling of hidden variables across different policies in the set \(\Omega\), where \(\Omega = \{\Omega_1, \Omega_2, \ldots, \Omega_n\}\), each represented by a multilayer perceptron (MLP) network~\cite{taud2018multilayer}. This pooling operation allows for information sharing even during execution, distinguishing our method from standard CTDE.

While autonomous vehicles are equipped with lidar sensors for local environment perception, they lack information about distant vehicles. By comparing these two approaches, we aim to evaluate the trade-offs in efficiency and safety between decentralized and centralized decision-making in autonomous vehicle scenarios. 

\subsection{Rule-Based Coordination}

As we refine the methods of distributed training to achieve optimal performance, it is equally crucial to align the reward functions closely with real-world scenarios. Effective cooperation among vehicles not only depends on individual performance but also on how well the system incentivizes safe and cooperative behavior. This leads us to explore rule-based coordination strategies that address common traffic challenges more realistically.


\paragraph{Safe Distance}The objective of this experiment is to explore the impact of maintaining safety distances in multi-agent environments. Maintaining an appropriate distance between vehicles is essential for safe driving operations.
In our experiment, the front 10 lidar points of the vehicle are designated as the sensing area, covering a 50-degree angle in front. 
\begin{figure}[h]
    \centering
    \includegraphics[width=\linewidth]{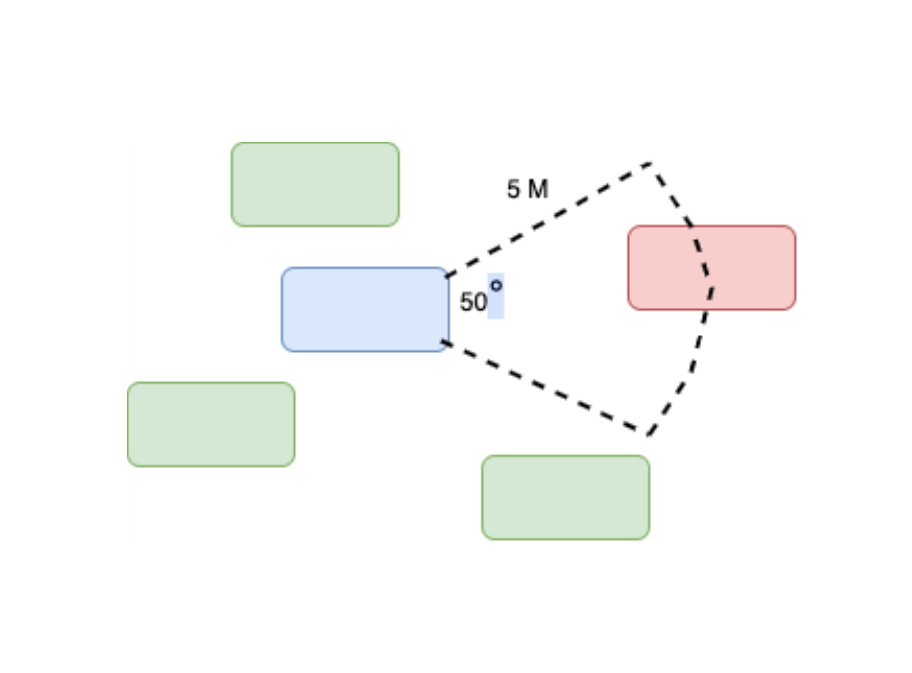}
    \caption{The blue vehicle is penalized for being too close to the red vehicle in front of it. However, surrounding green vehicles, due to either sufficient distance or not being directly ahead of the blue vehicle, do not trigger a penalty for the blue vehicle.}
    \label{fig:safe_distance}
\end{figure}
When a vehicle ahead is detected within a distance of less than 5 meters, a penalty is imposed on the vehicle as illustrated in Figure \ref{fig:safe_distance}. The safe distance penalty is calculated using the formula:
\begin{equation}
R_{\text{sd},i} = -0.5 \times \left(\frac{5m - d_{i,j}(t)}{5m}\right)
\end{equation}
where \(d_{i,j}(t)\) represents the distance between vehicle \(i\) and vehicle \(j\) at time \(t\). This setup aims to simulate the importance of maintaining safe distances in real driving and encourages agents to learn to avoid dangerously close behaviors through negative incentives. Ultimately, the calculated penalty \( R_\text{sd} \) is added into the existing local reward function \( R_{local} \), enhancing the model's ability to train agents on the significance of maintaining safe distances.

\paragraph{Right-of-way Responsibility}In our simulated driving environment, a common issue encountered is the dual penalization of vehicles involved in collisions, irrespective of the actual traffic fault. This approach often contradicts real-world traffic regulations, where typically only one party is deemed primarily responsible for the incident. This discrepancy between simulated and real-world scenarios prompted us to reevaluate the fairness and realism of our penalty system.

In reality, the right-of-way rules dictate that not all vehicles involved in an accident should bear equal responsibility~\cite{li2021three}. This principle is well established in traffic law but less commonly represented accurately in reinforcement learning simulations. To address this, we implemented a simple rule-based heuristic to determine the primary responsible vehicle in multi-vehicle collisions, defined as \( Z(i, j) \).
\begin{equation}
R_{\text{rc},i} = \begin{cases} 
2 \cdot R_{\text{collsion}, i} & \text{if } Z(i, j)  \\
0 & \text{otherwise}
\end{cases}
\end{equation}
This method leverages known traffic rules and right-of-way principles to assign fault more accurately, thereby aligning our simulation more closely with real-world legal frameworks. This adjustment not only enhances the realism of our model but also allows for more nuanced assessments of autonomous driving policies under various traffic conditions.

\begin{table}
\caption{Performance Comparison between CTDE and CTCE Approaches}
\label{table:ctde_vs_ctce}
\begin{center}
\begin{tabular}{lrr}
\hline
                                      &   CTDE &   CTCE \\
\hline
 success                             &      40    &  40    \\
 out\_of\_road                         &      14.39 &  \textbf{10.32} \\
 crash\_vehicle                       &      31.43 &  \textbf{28.37} \\
 velocity\_mean                       &       \textbf{4.11} &   3.64 \\
 velocity\_mean\_in\_conflict\_zone      &       \textbf{2.47} &   2.11 \\
 acceleration                        &       0.6  &  \textbf{0.7}  \\
 acceleration\_in\_conflict\_zone       &       0.43 &   \textbf{0.49} \\
 arrive\_steps                        &     \textbf{277.96} & 313.76 \\
 episode\_steps                       &     \textbf{478.92} & 534.72 \\
 mean\_conflict\_zone\_num              &       5.73 &   \textbf{6.33} \\
 max\_conflict\_zone\_num               &      12.12 &  \textbf{13.2}  \\
 front\_end\_distance                  &       0.18 &   0.18 \\
 limited\_lidar                       &       \textbf{0.56} &   0.55 \\
 limited\_lidar\_in\_conflict\_zone      &       0.43 &   0.41 \\
 front\_end\_distance\_in\_conflict\_zone &       \textbf{0.12} &   0.11 \\
 pair\_distance                       &      \textbf{37.76} &  37.01 \\
\hline
\end{tabular}
\end{center}
\end{table}

\section{Experiments And Results}

In this section, we present the evaluations of our experiments. Our training lasted 24 hours, and we choose the best model to present. There are many angles in the SVO experiment, so we only trained for 12 hours and took the best model. All models were evaluated 100 times to ensure the accuracy of the experiments. Finally, we discussed the impact of different conditions on safety and efficiency.


\subsection{Performance Indicators}

To understand the trade-off between safety and efficiency, we must consider four main indicators. For safety, these indicators include the number of collisions and the number of vehicles out of the road. For efficiency, we focus on the total number of vehicles that successfully arrive at the destination and the average speed of the vehicles. A detailed addition is that, since vehicles do not disappear after collisions in our simulations setting, it is possible for multiple collisions to occur between the same vehicles in a short period. We also monitor situations such as obstructions in the vehicle's field of vision, distance to the front vehicle, and vehicle density. In this paper, due to environmental constraints, we use lidar obstruction to represent visual obstructions. Additionally, we pay close attention to vehicle performance in the central area, or the conflict zone, where most collisions occur. For efficiency, we also focus on the average number of steps to reach all vehicles and the steps taken by the last vehicle to arrive, corresponding to the episode steps in reinforcement learning.


\subsection{Centralized Policy}

CTDE and CTCE share the same objective function, aiming to balance safety and efficiency in autonomous vehicle coordination. As demonstrated in Table~\ref{table:ctde_vs_ctce}, while CTCE exhibits improved safety with fewer out-of-road incidents and slightly fewer crashes, it does so at the expense of reduced velocity and extended episode steps. This suggests that CTCE may adopt a more cautious strategy, prioritizing safety over speed. Conversely, CTDE tends to prioritize efficiency, evidenced by higher velocities and shorter episode completions, though this approach comes with a slight increase in safety risks, as indicated by more frequent out-of-road incidents and crashes. The increased conflict zone numbers in CTCE suggest that it may be more effective at coordinating multiple vehicles through intersections simultaneously, possibly reflecting a more sophisticated approach to traffic management, despite a slightly slower velocity.

\subsection{Rule-Based Reward Function Policies}

Table \ref{table:safety_result} presents the outcomes of experiments using rule-based rewards, comparing the baseline model with models that include penalties for not maintaining safe distances and not adhering to right-of-way rules, along with their combined effects.

Implementing a safe distance penalty alone reduced crash incidents from 31.43 to 14.62, demonstrating the effectiveness of embedding safety-aware behaviors through negative reinforcement. When combined with right-of-way rules, the system's performance improved further, reducing the crash rate to 12.61 and out-of-road incidents to 13.48. The mean conflict zone number decreased to 3.91, indicating enhanced navigational safety and efficiency. Notably, these safety improvements came with only a slight reduction in vehicle efficiency.

These results show that rule-based reward functions can significantly influence autonomous vehicle behavior, promoting safer and more efficient driving practices. While not necessarily consistent with all real-world traffic rules, this approach demonstrates potential for improving traffic flow and safety outcomes in simulated environments.



\section{CONCLUSIONS}

This paper has explored various strategies to enhance the safety and efficiency of autonomous vehicle systems through the implementation of a novel distributed training framework OPTIMA. The results have demonstrated that the integration of these strategies can significantly influence vehicle behavior, promoting safer and more efficient traffic management in simulated environments. This approach has led to notable improvements in both the safety and efficiency of CAVs in challenging traffic situations, particularly in multi-agent environments like intersections. Moving forward, while this study has focused on multi-agent systems using homogeneous policies, future work could investigate the integration of heterogeneous policy strategies. This approach would explore how different policy strategies interact and potentially enhance the overall safety and efficiency of CAVs. Such studies could provide deeper insights into the dynamic interactions within multi-agent systems and lead to more robust autonomous transportation solutions.

\clearpage






\bibliographystyle{IEEEtran}
\bibliography{references}
\end{document}